\documentstyle[emulateapj,epsf]{article}

\def\ga{\gtrsim}
\def\la{\lesssim}
\def\etal{{\it et al.,\ }}
\def\eg{{e.g.,\ }}
\def\ie{{i.e.,\ }}
\def\rmp{{Rev. of Modern Phys.\ }}
\begin{document}

\title{Critical self-organization of astrophysical shocks} 
\author{M. A. Malkov,  P. H. Diamond}
\affil{ University of California at San Diego, 9500 Gilman Dr,
La Jolla, CA 92093-0319, USA  \\
\it mmalkov@physics.ucsd.edu; pdiamond@physics.ucsd.edu \rm} 
\and 
\author{H. J. V\"olk}
\affil{Max-Planck Institut f\"ur Kernphysik, D-69029, Heidelberg, Germany}

\begin{abstract}
There are two distinct regimes of the first order Fermi acceleration
at shocks.  The first is a linear (test particle) regime in which most
of the shock energy goes into thermal and bulk motion of the plasma.
The second is an efficient regime when it goes into accelerated
particles.  Although the transition region between them is narrow, we
identify the factors that drive the system to a {\it self-organized
critical state} between those two.  Using an analytic solution, we
determine this critical state and calculate the spectra and maximum
energy of accelerated particles.
\end{abstract}

 
\section{Introduction}\label{intr}
It is now generally recognized that most of the observed gamma
radiation derives in one way or another from accelerated particles.
Radio and x-ray spectral components from a variety of astrophysical
objects are believed to have a similar origin.  High energy neutrinos,
whose detection is on the program for the future and existing
water/ice detectors, must also be related to the ultra-high energy
cosmic rays (UHECR).  Their origin is, in turn, a mystery and the huge
Auger detector complex is now being built to elucidate it
(\cite{bla99,cron99}).

There has been essential progress in our understanding of how
accelerated particles produce the radiation detected.  The models
concentrate on the synchrotron and inverse Compton emission for
electrons and on the $\gamma$ ray and neutrino production in $pp $ and
$p\gamma$ reactions for protons. However, the primary spectrum of
accelerated particles remains a stumbling block  making predictions
of otherwise similar models so different (\eg \cite{mpr99,wb99}).

The ``standard'' mechanism of particle acceleration, capable of
producing nonthermal power-law spectra extending over many decades in
energy is the I-order Fermi or diffusive shock acceleration.  It was
originally suggested to explain the origin of galactic cosmic rays
(CRs).  For the purposes of the high-energy radiation and UHECR it is
usually adopted as an axiom, and mostly only in its simplest, test
particle (TP) or linear realization.  In particular, it is assumed
that any strong nonrelativistic shock routinely produces a $E^{-2} $
spectrum of protons and/or electrons.  In fact this spectrum arises
from a simple formula $F\sim E^{-(r+2)/(r-1)}$ where $r$ is the shock
compression (for $r=4$).  However, it is valid only if the shock
thickness is much smaller than the particle mean free path.  This, in
turn, is true only if the energy content of accelerated particles is
small compared to the shock energy (inefficient acceleration) so that
the shock structure is maintained by the thermal, not by the high
energy particles.  Otherwise, the accelerated particles create the
shock structure on their own and if so, then obviously on a scale that
is larger or of the order of their mean free path, thus making the
above formula invalid.  Therefore, the TP regime requires a very low
CR number density (the rate of injection $\nu$ into the acceleration
process), which appears to be impossible in the parameter range of
interest.  It has been inferred from observations (\eg \cite{lee82}),
simulations (\eg \cite{be95}), and theory (\eg \cite{m98}) that the CR
number density $n_{\rm CR}$ at a strong shock must be $\sim 10^{-3}$
of the background density $n_1 $ upstream.  It is important to
emphasize here that when the actual injection rate $\nu$ exceeds the
critical value (denote it $\nu_2$), the test particle ($E^{-2}$)
solution simply does not exist. Simple measures, such as calculating
corrections, are intrinsically inadequate.  What happens is that two
other solutions with considerably higher efficiencies branch off at a
somewhat lower injection rate $\nu÷_1 <
\nu_2$, one of which disappears again at $\nu÷=\nu÷_2$, together with
the test particle solution.

Thus, it seems to be difficult to put an accelerating shock into a
regime in which the CR energy production rate (acceleration
efficiency) could be gradually adjusted by changing parameters.  It is
either too low (TP regime) or it is close to unity. Note, that this
situation is quite suggestive of that occurring in phase transitions
or bifurcations.

Generally, neither of those extreme regimes provide an adequate
description of particle spectra and related emission.  Nevertheless,
we argue in this {\it Letter} that despite this apparent lack of
regulation ability, shocks must be still capable of {\it
self-regulation} and {\it self-organization}.  The transition region
between the two acceleration regimes (critical region) is very narrow
in control parameters like $\nu $. On the other hand, the
self-regulation can work efficiently only when the parameters are
within this region.  This requirement determines them, and also
resolves the question of the mechanism of self-regulation.  The above
consideration is similar to the concept of {\it self-organized
criticality} (SOC) (\eg
\cite{bak87}, \cite{hwa},
\cite{dh95}).
\section{Formulation of the problem}\label{form}
We use the diffusion-convection equation (\eg \cite{dru83}) for
describing the distribution of high energy particles (CRs).  We assume
that the gaseous discontinuity (also called the subshock) is located
at $x=0$ and the shock propagates in the positive $x$- direction.
Thus, the flow velocity in the shock frame can be represented as
$V(x)=-u(x)$ where the (positive) flow speed $u(x)$ jumps from
$u_2\equiv u(0-)$ downstream to $u_0\equiv u(0+) > u_2 $ across the
subshock and then gradually increases up to $u_1\equiv u(+\infty ) \ge
u_0$.  In a steady state the equation reads
\begin{equation}
		u\frac{\partial f}{\partial x} +\kappa(p)
\frac{\partial^2 f}{\partial x^2} = \frac{1}{3} \frac{du}{dx}
p\frac{\partial f}{\partial p}, \protect\label{c:d}
\end{equation}
where $f(x,p)$ is the isotropic (in the local fluid frame) part of the
particle distribution.  This is assumed to vanish far upstream ($f\to
0, \, x\to \infty$), while the only bounded solution downstream is
obviously $f(x,p)=f_0(p)\equiv f(0,p)$.  The most plausible assumption
about the cosmic ray diffusivity $\kappa(p)$ is that of the Bohm type,
\ie $\kappa(p)=K p^2/\sqrt{1+p^2}$ (the particle momentum $p$
is normalized to $mc$).  In other words $\kappa$ scales as the
gyroradius, $\kappa \sim r_{\rm g}(p)$.  The reference diffusivity $K$
depends on the $\delta B/B$ level of the MHD turbulence that scatters
the particles in pitch angle.  The minimum value for $K$ would be $K
\sim mc^3/eB$ if $\delta B \sim B$. Note that this plain
parameterization of this important quantity is perhaps the most
serious incompleteness of the theory which will be discussed later.

To include the backreaction of accelerated particles on the plasma
flow three further equations are needed.  First, one simply writes the
conservation of the momentum flux in the smooth part of the shock
transition ($x > 0$, the so-called CR-precursor)
\begin{equation}
		P_{\rm c}+\rho u^2   = \rho_1 u_1^2 , \quad x>0
	\label{mom:c}
\end{equation}
where $P_{\rm c}$ is the pressure of the high energy particles
\begin{equation}
	P_{\rm c}(x)= \frac{4\pi}{3} mc^2 \int_{p_0}^{p_1}\frac{p^4 
dp}{\sqrt{p^2+1}} f(p,x)
	\label{P_c}
\end{equation}
It is assumed here that there are no particles with momenta $p > p_1$
(they leave the shock vicinity because there are no MHD waves with
sufficiently long wave length $\lambda $, since the cyclotron
resonance requires $p \sim \lambda$).  The momentum region $0 < p <
p_0$ cannot be described by equation (\ref{c:d}) and the behavior of $f(p)$
at $p\sim p_0$ is described by the injection parameters $p_0$ and
$f(p_0)$ (\cite{m97a}, [M97]).  The plasma density $\rho (x)$ can be
eliminated from equation (\ref{mom:c}) by using the continuity equation
$\rho u =\rho_1 u_1$.  Finally, the subshock strength $r_{\rm s}$ can
be expressed through the Mach number $M $ at $x=\infty$ (\eg
\cite{ll:hd})
\begin{equation}
		r_{\rm s} \equiv
	\frac{u_0}{u_2}=\frac{\gamma+1}{\gamma-1+ 2R^{\gamma+1}M^{-2}}
	\label{c:r}
\end{equation} 
where the precursor compression $R\equiv u_1/u_0$ and
$\gamma$ is the adiabatic index of the plasma.

The system of equations (\ref{c:d},\ref{mom:c},\ref{c:r}) describes in
a self-consistent manner the particle spectrum and the flow
structure. An efficient way to solve it is to reduce this system to one
integral equation (M97).  A key dependent variable is an integral
transform of the flow profile $u(x)$ with a kernel suggested by an
asymptotic solution of the system (\ref{c:d})-(\ref{mom:c}) which has the form
$$f(x,p)=f_0(p) \exp \left[-\frac{q}{3\kappa}\Psi\right]$$ 
where $$\Psi=\int_{0}^{x}u(x') d x'$$ and the spectral index
downstream $q(p)=-d \ln f_0/d \ln p$.
The
integral transform is as follows
\begin{equation}
U(p)=\frac{1}{u_1}\int_{0-}^{\infty} \exp
\left[-\frac{q(p)}{3\kappa(p) }\Psi\right]du(\Psi) \label{U}
\end{equation}
and it is related to $q(p)$ through the
following formula
\begin{equation}
	q(p)=\frac{d \ln U}{d \ln p} +\frac{3}{r_{\rm s} R U(p)}+3
	\label{q(p)}
\end{equation}
Thus, once $U(p) $ is found both the flow profile and the particle
distribution can be determined by inverting transform (\ref{U}) and
integrating equation (\ref{q(p)}).  Now, using the linearity of
equation (\ref{mom:c}) ($\rho u=const $), we derive the integral
equation for $U$ by applying the    transformation (\ref{U}) to
the $x-$ derivative of equation (\ref{mom:c}) (M97). The result reads
\begin{eqnarray}
	U(t) &= & \frac{r_{\rm s}-1}{R r_{\rm s}}+
	\frac{\nu}{K p_0}\int_{t_0}^{t_1}d t'
	\left[\frac{1}{\kappa(t')}+\frac{q(t')}
	{\kappa(t)q(t)}\right]^{-1}
	\nonumber \\
	       &\times & 
	\frac{U(t_0)}{U(t')} \exp \left[-\frac{3}{R r_{\rm
	s}}\int_{t_0}^{t'}\frac{d t''}{U(t'')}\right] 
	\label{int:eq}
\end{eqnarray}
where $t=\ln p$, $t_{0,1}=\ln p_{0,1}$. Here the injection parameter
	\begin{equation} \nu÷=\frac{4\pi}{3}\frac{mc^{2}}{\rho_1
	u_1^2} p_0^4 f_0(p_0) \label{nu:def} \end{equation} is related
	to $R$ by means of the following equation
\begin{eqnarray}
	\nu÷&=& K p_0
		\left(1-R^{-1}\right) \nonumber \\
		&\times & \left\{\int_{t_0}^{t_1}\kappa(t) d t
		\frac{U(t_0)}{U(t)}\exp \left[-\frac{3}{R r_{\rm
		s}}\int_{t_0}^{t}\frac{d t'}{U(t')}\right]\right\}^{-1}
	\label{nu}
\end{eqnarray}
The equations (\ref{c:r},\ref{int:eq},\ref{nu}) form a closed system
that can be easily solved numerically. We analyze the results in the
next section.
\section{Mechanisms of critical self-organization}\label{soc}
The critical nature of this acceleration process is best seen in
variables $R,\nu$.  The quantity $R-1$ is a measure of shock
modification produced by CRs, in fact $(R-1)/R=P_{\rm c}(0)/\rho_1 u_1^2$
(eq.[\ref{mom:c}]) and may be regarded as an order parameter.  The
injection rate $\nu$ characterizes the CR density at the shock front
and can be tentatively treated as a control parameter.  It is
convenient to plot the function $\nu÷(R)$ instead of $R(\nu) $ (using
equation [\ref{nu}]), since $R(\nu÷)$ is not always a single-valued
function, Fig.  \ref{NuOfR:fig}.

The injection rate $\nu $ at the subshock should be calculated given
$r_{\rm s}(R)$ (M97) with the self-consistent determination of the
flow compression $R$ on the basis of the $R(\nu÷)$ dependence
obtained.  However, in view of its critical character, this solution
can be physically meaningful only in regimes far from criticality, \ie
when $R\approx 1$ (test particle regime) or $R \gg 1 $ (efficient
acceleration).  But, it is difficult to see how this system could
stably evolve remaining in one of these two regimes.  Indeed, if $\nu$
is subcritical it will inevitably become supercritical when $p_1$ is
sufficiently high.  Once it happened, however, the strong subshock
reduction (equation [\ref{c:r}]) will reduce $\nu $ and drive the
system back to the critical regime, Fig. \ref{bif:fig}.

The maximum momentum $p_1$ is subject to self-regulation as well.
Indeed, when $R \gg 1$, the generation and propagation of Alfven waves
is characterized by strong inclination of the characteristics of wave
transport equation towards larger wavenumbers $k $ on the
$k -x$ plane due to wave compression.  Thus, considering
particles with $p \la p_1$ inside the precursor, one sees that they
are in resonance with waves that must have been excited by particles
with $p > p_1$ further upstream but, there are no particles with $p >
p_1$.  Therefore, the required waves can be excited only locally by
the same particles with $p \la p_1$ which substantially diminishes the
amplitude of waves that are in resonance with particles from the
interval $ p_1/R < p < p_1$. (The left inequality arises from the
resonance condition $kcp
\approx eB/mc $ and the frequency conservation along the
characteristics $ku(x) = const$). This will worsen the confinement of
these particles to the shock front. The quantitative study of this
process is the subject of current research.  What can be inferred from
Fig.  \ref{bif:fig} now, is that the decrease of $p_1$ straightens and
rise the curve $\nu÷(R)$, so that it returns to the monotonic
behaviour.  However once the actual injection becomes subcritical (and
thus $R \to 1$) then $p_1 $ will grow again restoring the two extrema
on the curve $\nu(R)$.

The above dilemma is quite typical for dynamical systems that are
close to criticality or marginal stability. A natural way to resolve
it consists in collapsing the extrema into an inflection point so that
a {\it self-organized critical} (SOC) acceleration regime is
established being determined by the conditions
$\nu÷'(R^*)=\nu÷''(R^*)=0$.  These conditions not only determine
unique critical values $R^*$ and   $\nu^*\equiv \nu(R^*)$  
but also yield the maximum momentum $p_1$ as a function of $M$, which
is shown in Fig.
\ref{maxm:fig}.
A few particle spectra that develop in the SOC states for different
Mach numbers are shown in Fig.  \ref{ind:fig}, along with the
asymptotic $M=\infty $ non SOC spectrum. The latter can be calculated
in a closed form (M97). Note that the hardening of the spectra in
about the last decade below the cut-off is entirely due to the abrupt
cut-off itself.
\section{Discussion}\label{disc}
The detailed microphysics behind the SOC is extremely complex and must
include the self-consistent turbulence evolution and particle
acceleration with their strong backreaction on the shock structure.
We have simplified it and argued that the most important dynamical
components, the bulk plasma flow and the high-energy particles, must
be in a balance that constitutes a certain equipartition of the shock
energy between the two.  This was done by identifying the factors that
prevent either of them from prevailing alone.

The above situation is similar to that in \eg a simple sandpile
paradigm of the SOC. It is impossible (in fact unnecessary) to
describe the individual grain dynamics, but it is clear that when the
critical macroscopic characteristic of the system (the slope of the
sandpile) becomes too steep due to the action of external factors,
like tilting of the entire system or addition of sand at the top, the
sandpile relaxes bringing the slope to its critical magnitude.
\section{Possible feedback from observations}\label{obs}
Perhaps the most significant observational aspect is the particle
spectrum.  Although the conversion of detected radiation spectra into
the primary particle spectra is ambiguous, in some cases it may be
compared with the theory.  The most striking prediction is that in
shocks with {\it very high maximum particle energy}, the spectra must
be harder than $q=4$ (or 2 in the normalization $f(E)dE$).  This is
because of a very low injection requirement for efficient acceleration
in such shocks (Fig.\ref{NuOfR:fig}).  If we (conservatively) set
$n_{\rm CR}/n_1
\sim 10^{-3}$, then $\nu \sim 10^{-3} cp_0/mu_1^2 $ which may
easily exceed $\nu_2 $ (local maximum on the $\nu(R)$ curve) already
for $p_1 \ga 10^4 - 10^5 $, putting the acceleration into a strongly
nonlinear regime.  This should have important observational
consequences.

First, the particle energy is concentrated at the upper cut-off
instead of being evenly distributed over the logarithmic energy bands
as in the test particle $E^{-2} $ solution.  This makes the upper
bounds on CR generated neutrino fluxes (see \eg \cite{bw99} and
\cite{mpr99} and references therein) rather ambiguous.  Indeed, 
the UHECR spectrum is normalized to the observed one at $E_{\rm norm}
\sim 10^{19}$ eV while being obscured by the galactic background at
the energies $E \la 10^{18}$ eV.  Thus, CR spectra harder than
$E^{-2}$ imply that the upper limit on the neutrino fluxes \eg derived
by
\cite{wb99} should be even lower than the $E^{-2}$ spectrum implies for 
the energies $E_\nu <5\cdot 10^{17}$ eV ($E_{\nu}/E_{\rm CR}\simeq
0.03 $ is a typical energy relation). According to the same logic, it
should be increased for higher energies.  Note that the upper cut-offs
in individual shocks contributing to the UHECR must be still much
higher than $E_{\rm norm}$ to validate our simplified handling of
particle losses.  On the other hand if the sources with $E_{\rm max} <
E_{\rm norm}$ contribute significantly, the measurements at $E_{\rm
norm} $ tell us nothing about their normalizations and the upper bound
on neutrino fluxes may be increased up to the level dictated by the CR
observations in the lower energy range, as suggested by
\cite{mpr99}. This scenario is supported by Fig. \ref{maxm:fig}, 
provided that there are many strong shocks in the ensemble.  The
observed CR steep power-law spectrum is then essentially a
superposition of flatter or even non-power-law spectra from individual
sources properly distributed in $E_{\rm max}$.

To summarize our conclusions, the main factor that should determine
the particle primary spectra, and thus the neutrino flux, is how the
accelerating shocks are distributed in cut-off momenta, which in a SOC
state means in Mach numbers. There is no universal spectral form for
individual shocks at the current state of the theory (except a not
quite representative case of $M\to
\infty $).
Therefore, one should understand particle losses mechanisms, since
they determine the shock structure and thus the spectra, directly and
through $E_{\rm max}$. These mechanisms are inseparable from the
dynamics of strong compressible MHD turbulence generated by those same
particles.  The further progress in its study will improve our
understanding of the acceleration process and related radiation.

\acknowledgments
This work was supported by U.S. DOE under Grant No. FG03-88ER53275.
We also acknowledge helpful discussions with F. A. Aharonian and
J. P. Rachen.





\plottwo{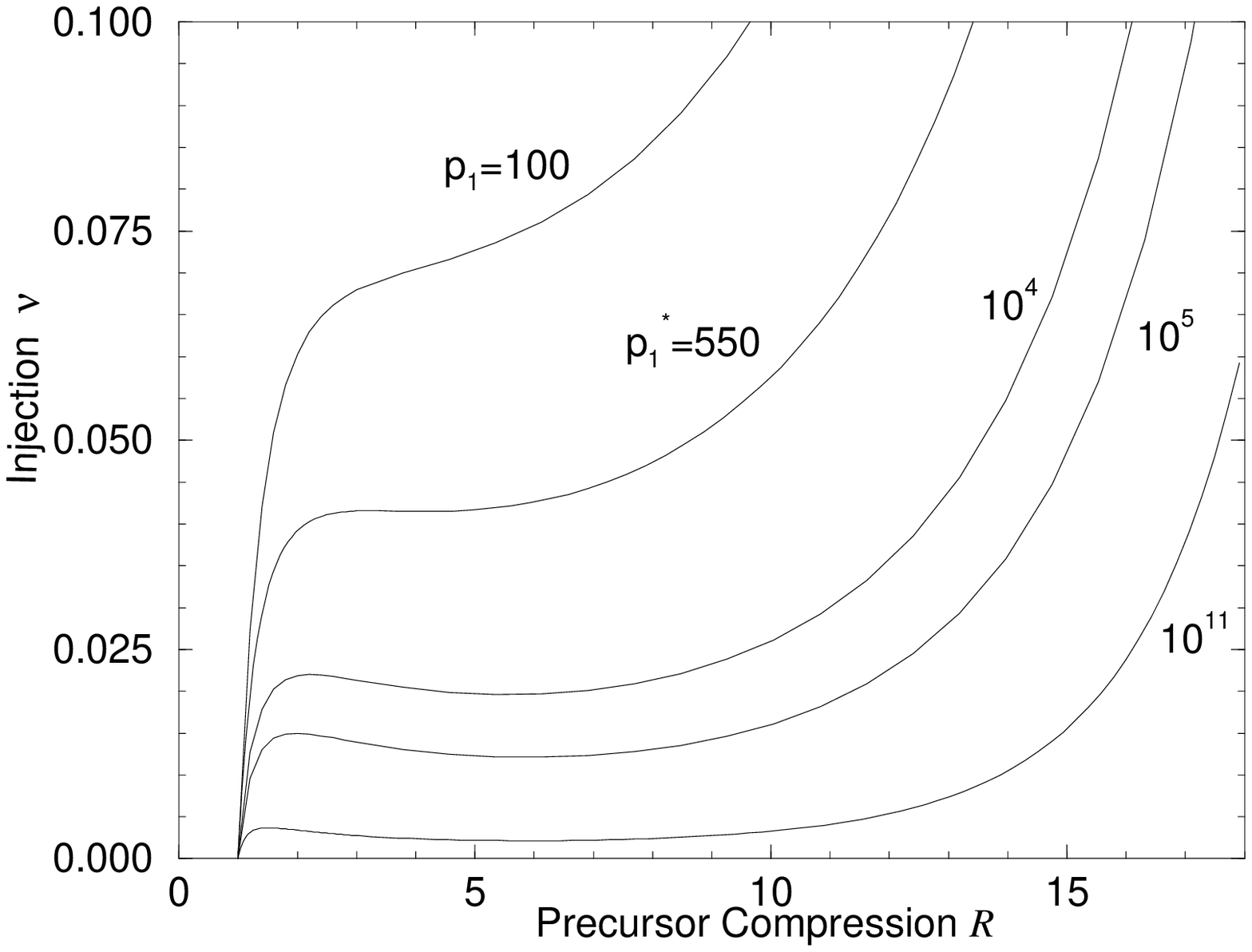}{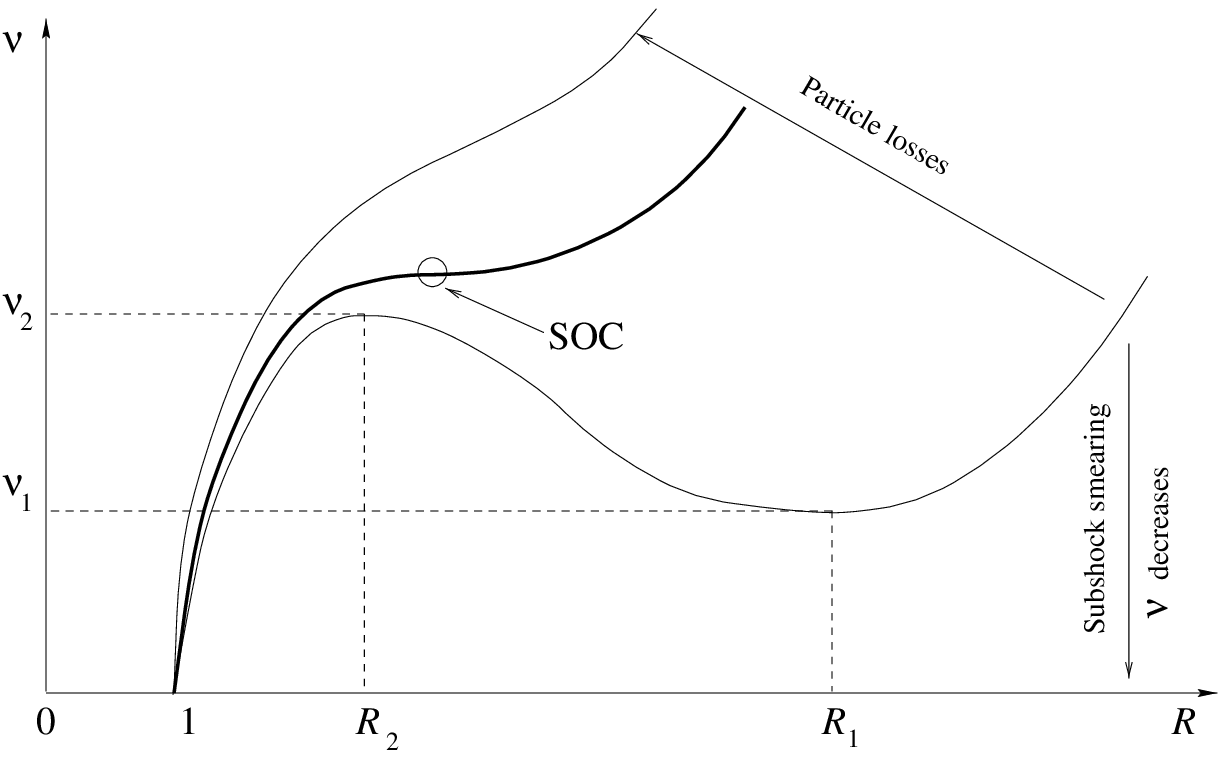}


	\figcaption[fig1.eps]{The nonlinear response of an
	accelerating shock (characterized by the precursor compression
	$R$) to the thermal injection $ \nu $ given in the form of the
	function $\nu(R) $ calculated for the fixed injection momentum
	$p_0=10^{-3} $, Mach number $M=150$ and for different
	$p_1=100; 550; 10^4; 10^5; 10^{11}$. The critical value (see
	text) $p_1^*=550$. The TP regime is limited to the region
	$R\simeq 1$. \protect\label{NuOfR:fig}}

	\figcaption[fig2.eps]{Bifurcation diagram corresponding to the
	set of response curves shown in fig. \ref{NuOfR:fig}. Since
	$\nu $ and $p_1$ are in reality dynamic rather than control
	parameters the response curve moves towards the bifurcation
	curve drawn with the heavy line.  It corresponds to the curve
	marked by $p_1^*=550$ in Fig. \ref{NuOfR:fig}.
	\label{bif:fig}}


\plottwo{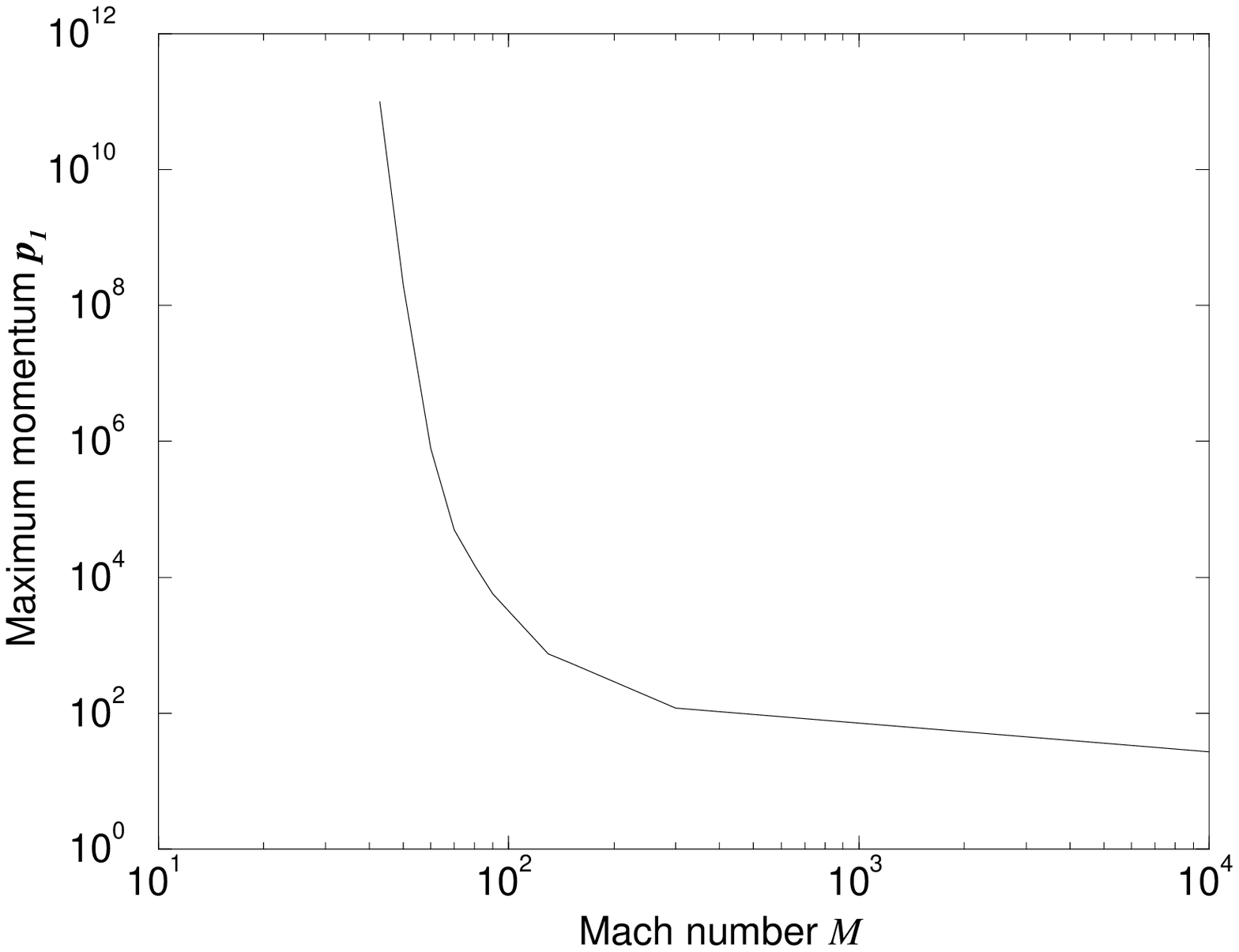}{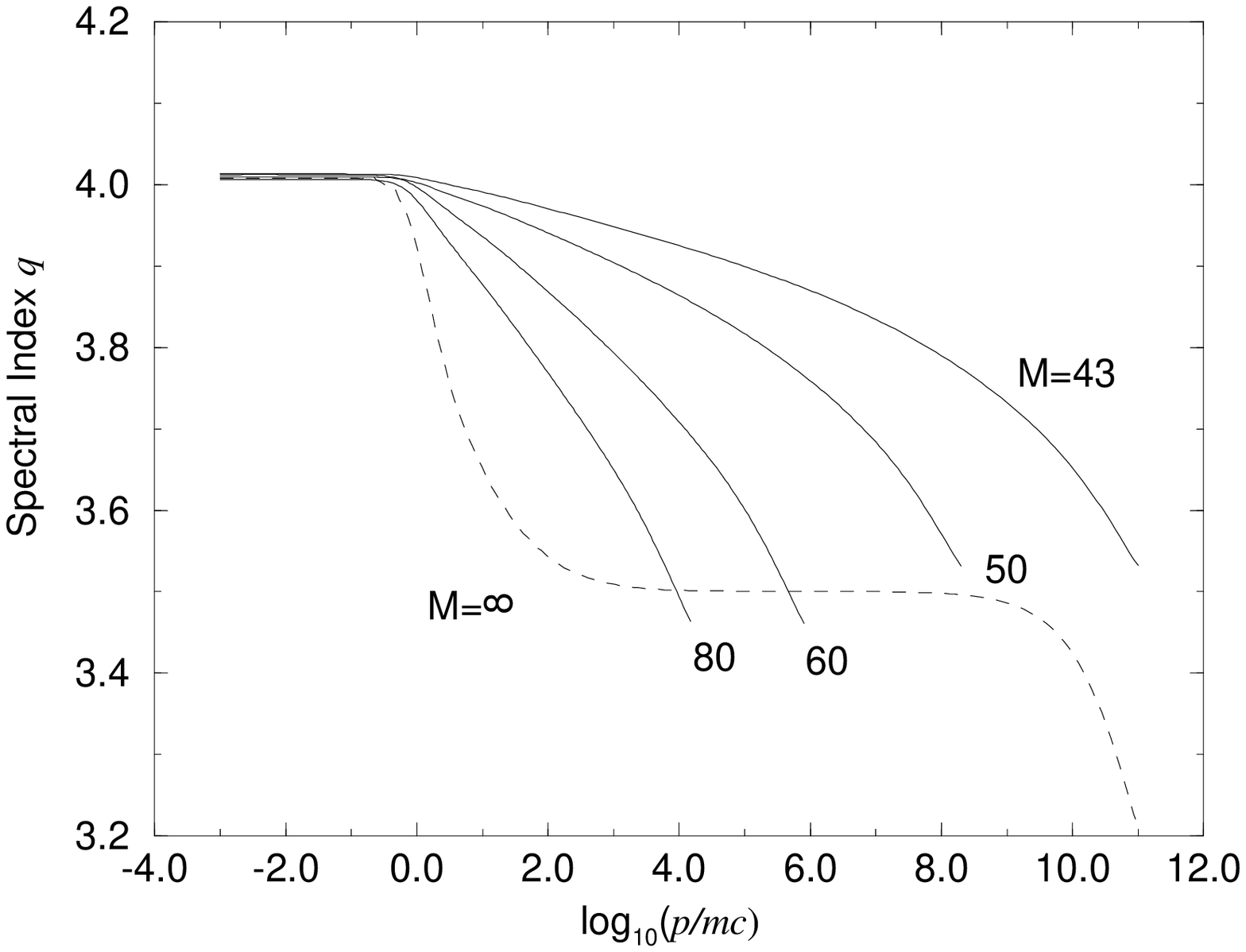}





	\figcaption[fig3.eps]{Maximum momentum $p_1$ versus Mach
	number $M$ calculated in the SOC states (see
	text).\label{maxm:fig}}

	\figcaption[fig4.eps]{The spectral indices in the SOC states
	shown for different Mach numbers and the corresponding maximum
	momenta from Fig. \ref{maxm:fig} (solid curves). For
	comparison, the dashed curve shows the asymptotic case
	$M=\infty, \, p_1=10^{11} $ and $\nu = 4\cdot 10^{-5} $. For
	larger, \eg SOC injection values this spectrum would have been
	cut in mildly relativistic region.
\label{ind:fig}}

\end{document}